# 10 W injection-locked single-frequency continuous-wave titanium:sapphire laser

**TETSUSHI TAKANO,**[1,*] **HISASHI OGAWA,**[1] **CHIAKI OHAE,** [2,3] **AND MASAYUKI KATSURAGAWA**[2,3]

[1] *NICHIA Corporation, 3-13-19, Moriya-Cho, Kanagawa-Ku, Yokohama, Kanagawa 221-0022, Japan*
[2] *Graduate School of Informatics and Engineering, University of Electro-Communications, 1-5-1 Chofugaoka, Chofu, Tokyo 182-8585, Japan*
[3] *Institute for Advanced Science, University of Electro-Communications, 1-5-1 Chofugaoka, Chofu, Tokyo 182-8585, Japan*
*\*tetsushi.takano@nichia.co.jp*

**Abstract:** High-power tunable lasers with good longitudinal and transverse modes are fundamental tools for exploring quantum physics. Here we report a high-power continuous-wave injection-locked titanium:sapphire laser with a low-loss cavity configuration, where only a laser crystal was installed in the laser cavity. Although the transverse mode was affected by a thermal lens formed in the laser crystal, the focal length of the thermal lens could be shifted via the temperature of the laser crystal holder or the pump power. As a result, we found a condition that 10 W single-frequency oscillation with a good transverse mode and a slope efficiency of 51% were achieved.

## 1. Introduction

Titanium:sapphire (Ti:sapphire) crystal has a broad emission wavelength range and high light damage threshold [1], which makes Ti:sapphire lasers ideal light sources for studying ultrafast physics [2], quantum metrology [3–5] and quantum information technology [6]. One of the important demands on light sources for such applications is the coexistence of high power and good transverse and longitudinal modes. Injection-locked Ti:sapphire lasers [7] have the advantage that their spectrum and power can be independently controlled. Dual-wavelength injection locking of Ti:sapphire lasers [8,9] has also been demonstrated and applied to Raman-resonant four-wave-mixing [10]. Since injection-locking makes it possible to omit wavelength-selective elements inside laser cavities, the laser efficiency can also be improved. In that context, a 6.5 W continuous-wave (CW) Ti:sapphire laser emitting at 852 nm with a linewidth of 1 kHz has been realized [11].

There has been growing demand for higher power lasers with narrow linewidth in the field of cold atom experiments like atom interferometers using multi-photon process [5], which needs a high power laser as possible with a spectral purity and a wavelength tunability to atomic resonance. Although the wavelength range of Ti:sapphire lasers covers many of the resonant wavelength of Alkali atoms, the output power of single-frequency injection locked Ti:sapphire lasers remained 5-6.5 W, even when 20 W class pump lasers were used [11,12]. Therefore, other 10 W-class lasers with a less tunable wavelength range have been chosen for above applications [13–15]. When developing a high-power Ti:sapphire laser with increased pump power, one crucial issue is how to manage the thermal lens effect [16], which results from variations in the refractive index in the laser crystal due to the thermo-optic effect caused by absorption of the pump beam [17]. In the case of a collimated pump beam for a pulsed Ti:sapphire laser, the property of the thermal lens and its influence on the laser performance has been reported [18]. However, to the best of our knowledge, injection-locked CW Ti:sapphire lasers with this effect have not been investigated well.

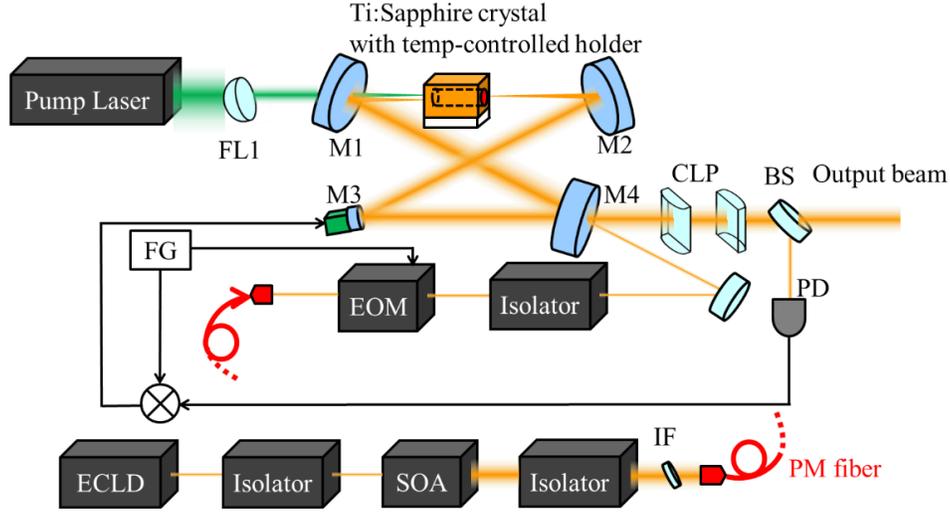

Fig. 1. Schematic illustration of our injection-locked titanium:sapphire (Ti:sapphire) laser. A laser cavity consisting of four mirrors M1-4 and Ti:sapphire crystal was pumped by a 532 nm laser, where the absorption coefficient of the Ti:sapphire crystal for the pump wavelength was about 1.80 /cm. 780 nm seed light, generated by a master oscillator power amplifier configuration of an external cavity laser diode (ECLD) and an optical semiconductor amplifier (SPA), was delivered through a polarization-maintaining (PM) fiber and injected into the laser cavity. The laser cavity was stabilized to the seed wavelength via a piezoelectric element connected to M3. An electro-optic modulator (EOM) and a photodetector (PD) were used to produce an error signal. A cylindrical lens pair (CLP) compensated for the astigmatism of the output beam. An interference filter (IF) rejected amplified spontaneous emission (ASE) of the SOA.

Here, we demonstrate a higher-power CW injection locked Ti:sapphire laser than ever reported. Since the Ti:sapphire cavity did not contain an isolator or a birefringent filter, which were included in the previous lasers [11,12], the round-trip loss decreased to 2%. By optimizing an output coupler and beam waist diameter of a pump beam for such configuration, slope efficiency was increased to 51% at a wavelength of 780 nm. Although a thermal lens affects the transverse mode in several conditions, its focal length could be shifted via the temperature of the laser crystal holder or the pump power. Finally, we found a condition that 10 W single-frequency oscillation with a good transverse mode and a slope efficiency of 51% were achieved.

## 2. Experimental setup

Figure 1 shows a schematic illustration of the developed injection-locked Ti:sapphire laser. The laser cavity consisted of two concave mirrors (M1 and M2) and two plane mirrors (M3 and M4). M1, M2, and M3 were provided with high-reflectivity (HR) coatings around a wavelength of 800 nm. A vertically-cut, rod-shaped Ti:sapphire crystal with anti-reflectivity (AR) coating was placed between M1 and M2. The dimension of the crystal was 5 mm in diameter and 20 mm in length. The absorption coefficient $\alpha_{532}$ of the Ti:sapphire crystal at 532 nm was 1.80 /cm and a figure of merit $\alpha_{532} / \alpha_{780}$ was > 250, where $\alpha_{780}$ means an absorption coefficient at 780 nm. The Ti:sapphire crystal was held in a temperature-controlled copper heat sink and an indium-sheet with a thickness of 0.05 mm was sandwiched between the crystal and the heat sink to improve thermal contact. A pump beam with a wavelength of 532 nm and a single transverse mode, which was obtained by a pump laser (Verdi G-20), was focused inside the Ti:sapphire crystal. The arm lengths of the Ti:sapphire cavity is described in Section 4. First, we designed the Ti:sapphire cavity as the Rayleigh length of the pump beam equal to a half of the absorption length $1 / \alpha_{532}$ [19], which corresponded to a beam waist radius of 16.4 µm, and then ended up choosing 15 µm as a result of an optimization of the Ti:sapphire cavity. Since any optical elements except for the Ti:sapphire crystal were not installed in the Ti:sapphire cavity, the round-trip loss of the cavity without the output coupler M4 was notably small and measured to be 2%. It was almost dominated by the absorption loss of the Ti:sapphire crystal of 1.4%. Since an optimum reflectivity was calculated to be 88.4% from the round-trip loss [20], we designed the reflectivity of the mirror M4 to be that value, and measured reflectivity was 88.7% in reality.

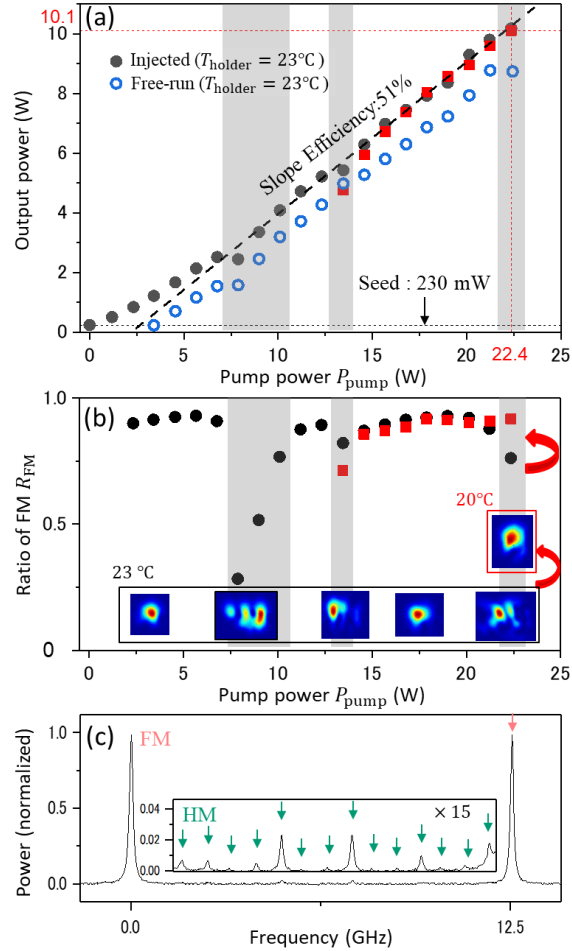

Fig. 2. Output power and transverse mode properties of the output beam. (a) Power of the output beam as a function of the pump power $P_{\text{pump}}$. Blue circles depict the power in the free-running case with the temperature of the laser holder $T_{\text{holder}}$ set to 23 °C, and black points depict the power in the injected case with $T_{\text{holder}}$ set to 23 °C. Black dashed line represents a linear fitting line of the injected case for $P_{\text{pump}} > 11$ W with $T_{\text{holder}}$ set to 23 °C, where the slope efficiency (SE) is 51 %. (b) Transverse mode properties of the output beam. Ratio $R_{\text{FM}}$ of the transverse fundamental mode power to that of all the transverse modes in the output beam in the injected case with $T_{\text{holder}}$ set to 23 °C, which was estimated using a Fabry-Pérot etalon. Images in the black rectangle show typical behaviors of the transverse mode in the injected case with $T_{\text{holder}}$ set to 23 °C. (c) Example of the transmission spectrum at $P_{\text{pump}} = 17.9$ W through the Fabry-Pérot etalon, which was designed to resolve transverse modes. Vertical axis is normalized to a fundamental peak height equal to unity. Two large peaks indicate the adjacent transverse fundamental mode (FM) peaks with a free spectral range of about 12.5 GHz, and many higher modes (HM) are located between them, as shown in the enlarged view. In the gray shaded regions in (a) and (b), the transverse mode degraded and the output power slightly decreased due to the thermal lens. Since the thermal lens was affected by $T_{\text{holder}}$, an output power of 10.1 W was achieved while $R_{\text{FM}}$ remained at more than 0.9 for a maximum pump power of 22.4 W by tuning $T_{\text{holder}}$ to 20 °C, as indicated by red squares in (a) and (b).

A focusing lens FL1 was slightly tilted to eliminate astigmatism of the pump beam, which was caused when the pump beam passed through the mirror M1. To control the wavelength of the Ti:sapphire laser, 780 nm seed laser light, emitted from a master-oscillator power-amplifier configuration consisting of an external cavity laser diode (ECLD) and a semiconductor optical amplifier (SOA) [21], was injected into the laser cavity. Here we chose the wavelength of 780 nm, which is a resonant wavelength of rubidium atoms. A narrow-band interference filter (IF) was placed after the amplifier to eliminate amplified spontaneous emission (ASE) of the amplifier [22]. The seed laser light was delivered via a polarization-maintaining (PM) fiber, phase-modulated by an electro-optic modulator (EOM), and then coupled to the cavity and stabilized using an error signal obtained by demodulating the signal from a photodetector (PD). After compensating for the astigmatism by using a cylindrical lens pair (CLP), the properties of the output beam from the mirror M4 were estimated.

## 3. Results

Figure 2(a) depicts the output power of the Ti:sapphire laser as a function of the pump power $P_{\text{pump}}$, which was measured in front of the laser cavity. The blue empty circles show the output power in the free-running case with the temperature of the Ti:sapphire crystal holder, $T_{\text{holder}}$, set to 23 °C, which was measured by adding the powers of both beams emitted from the mirror M4 in two directions. In the injected case, on the other hand, the output light was converged in one direction, since the seed laser light was injected into the laser cavity and stimulated emission due to the seed laser light dominated the round-trip light. The black points indicate the output power in the injected case with $T_{\text{holder}}$ set to 23 °C, the $y$-intercept of which shows the reflected seed power at the mirror M4 was about 230 mW with locking of the cavity. The round-trip seed beam lowered the lasing threshold [23] in the injected case to $P_{\text{pump}} \sim 1$ W from that in the free-running case, which was about 3 W. Above the threshold pump power, the slope efficiency gradually increased at $P_{\text{pump}} < 7$ W and became almost constant at $P_{\text{pump}} > 11$ W. The black dashed line indicates the result of a linear fitting for $P_{\text{pump}} > 11$ W, the slope efficiency of which was 51%.

Figure 2(b) depicts the ratio, $R_{\text{FM}}$, of the transverse fundamental mode power to that of all the transverse modes in the output beam, which was estimated from a transmission spectrum obtained through a Fabry-Pérot etalon, designed to resolve the transverse modes. Figure 2(c) shows an example of the transmission spectrum at $P_{\text{pump}} = 17.9$ W, where two large peaks indicate the adjacent transverse fundamental mode (FM) peaks, and many higher modes (HM) are located between them, as shown in the enlarged view. The black circles in Fig. 2(b) indicate $R_{\text{FM}}$ in the injected case with $T_{\text{holder}}$ set to 23 °C. The images in the black rectangle show typical behaviors of the transverse modes in the injected case, where the horizontal positions of the images indicate the corresponding pump power. Except for the gray shaded regions, $R_{\text{FM}}$ was at least 0.9, which corresponded to a $M^2$ value of 1.1, and a single Gaussian-like transverse mode was observed; however, in the gray regions the transverse mode degraded and the output power slightly decreased. As discussed in Section 4, we consider that this is because a thermal lens formed in the Ti:sapphire crystal affected the cavity parameters, causing the transverse modes of the cavity to become degenerate, which degraded the output beam.

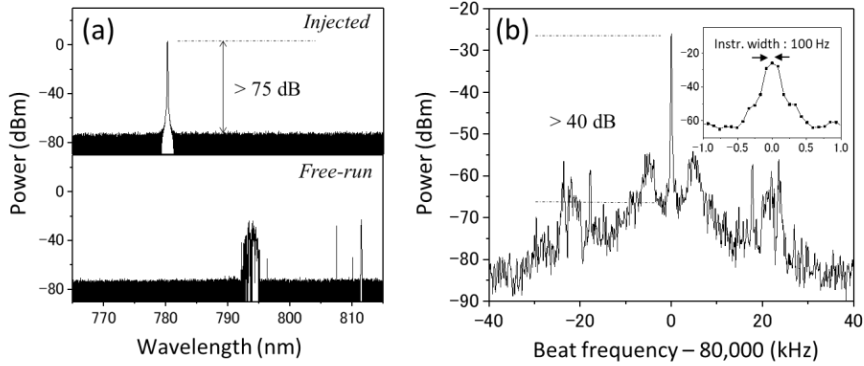

Fig. 3. Spectral properties of the output beam with $P_{pump}$ = 22.4 W and $T_{holder}$ = 20 °C. (a) Spectrum of the output beam measured by an optical spectral analyzer. The upper panel depicts the injected spectrum, which followed the seed laser with a single longitudinal mode, whereas the free-running spectrum shown by the lower panel was broad and unstable. The noise floor of the upper panel was determined by the instrumental limit, and was at least 75 dB lower than the spectral peak. No ASE spectrum was observed. (b) Narrow range spectral properties estimated from a beat signal between the output beam and the frequency-shifted seed beam. The central peak was clearly distinguished at more than 40 dB, and its linewidth was limited by the instrumental linewidth of 100 Hz, which indicates that the output beam reflected the same spectral properties as the seed laser without additional frequency noise, as far as we observed.

This kind of degradation due to a thermal lens is known to be a common problem in a high power laser. A heat capacitive active mirror has been proposed to solve this problem [24], but not realized yet. We found that this degradation is avoidable by slightly changing the temperature of the Ti:sapphire crystal holder $T_{holder}$. The red squares in Fig. 2(a) and (b) denote the behavior of the output beam in the injected case with $T_{holder}$ set to 20 °C, where an output power of 10.1 W was achieved while keeping $R_{FM}$ at more than 0.9 for a maximum pump power of 22.4 W.

Figure 3 depicts the longitudinal mode properties of the output beam. Fig. 3(a) shows a spectrum of the output beam with $P_{pump}$ = 22.4 W and $T_{holder}$ = 20 °C measured by an optical spectrum analyzer. The free-running spectrum was broad and unstable, as shown in the lower panel, whereas the injected one, shown in the upper panel, followed the single-mode spectrum of the seed laser at a wavelength of 780 nm. The noise floor of the upper panel was determined by the instrumental limit, which was about 75 dB lower than the spectral peak. No ASE spectrum was observed. The spectral properties in a narrower range were estimated from the beat signal between the output beam and the seed beam, frequency-shifted by an acousto-optic modulator. Figure 3(b) depicts the beat signal with an offset frequency of 80 MHz measured by an electric spectrum analyzer. The central peak was clearly distinguished at more than 40 dB, and its linewidth was limited by the instrumental linewidth of 100 Hz, which indicates that the output beam reflects the same spectral properties of the seed laser without additional frequency noise, as far as we observed.

### 4. Discussion

Although further experiments may be required to elucidate a complete mechanism of the results shown in Fig. 2, here we discuss a possible mechanism of the transverse mode degradation by means of ray transfer matrix analysis [25]. Figure 4(a) depicts the configuration of the laser cavity in detail, where we simply assume that one thermal lens is formed at the focal point of the pump beam in the Ti:sapphire crystal. The round-trip ABCD matrix of the laser cavity is calculated as

$$\begin{bmatrix} A & B \\ C & D \end{bmatrix}_{t,s} = M_{t,s}(R,\phi)S(L_1)S(n_c l_1)F(f_{\text{thermal}})S(n_c l_2)S(L_2)M_{t,s}(R,\phi)S(L_3), \tag{1}$$

where denotes an ABCD matrix for the concave mirrors M1 and M2 with a radius of curvature of $R = 100$ mm and an incident angle of $\phi = 18$ degree, $S(L_{1(2)})$ denotes that for the optical path between M1 (M2) and the incident (outgoing) facet of the Ti:sapphire crystal, with a length of $L_1 = 51$ mm ($L_2 = 45$ mm), $S(n_c l_{1(2)})$ denotes that for the optical path in the Ti:sapphire crystal between the thermal lens and the incident (outgoing) facet, with a length of $l_1 = 5$ mm ($l_2 = 15$ mm) and a refractive index of the Ti:sapphire crystal $n_c$ of 1.76, $S(L_3)$ denotes that for the other optical path with a length of $L_3 = 394$ mm, $F(f_{\text{thermal}})$ denotes that for the thermal lens with the focal length $f_{\text{thermal}}$, and the subscripts t and s denote the tangential plane and sagittal plane, respectively. The round-trip Gouy phase shift [26] is then calculated as

$$\varsigma_{t,s} = \arccos\left(\frac{A_{t,s} + D_{t,s}}{2}\right), \tag{2}$$

where $\varsigma_{t,s}$ is taken as $0 < \varsigma_{t,s} < \pi$ for $B_{t,s} > 0$ and $\pi < \varsigma_{t,s} < 2\pi$ for $B_{t,s} < 0$ [25]. Figure 4(b) depicts $\varsigma_{t,s}$ as a function of $f_{\text{thermal}}$. The lower limit of $f_{\text{thermal}}$ was calculated to be several mm according to the theoretical model based on the collinear geometry [16], where a broad collinear pump beam with a same intensity as our focused pump beam was applied to the laser crystal. Since an absorbed thermal quantity in our experiment was much smaller than that in the collinear geometry, we consider that a possible range of $f_{\text{thermal}}$ was more than several mm, which is shown by a gray shaded region in Fig. 4(b). The black and red curves indicate $\varsigma_t$ and $\varsigma_s$, respectively. When $\varsigma_{t,s}$ approaches $2\pi \times n/m$ where $m$ and $n$ are integers, the $m$-th order transverse mode in the Ti:sapphire cavity becomes degenerate with the fundamental mode [27]. Possible pairs of $m$ and $n$ with $m < 9$ were indicated by the blue dashed lines in Fig. 4(b). Since the transverse mode is split into the tangential direction (horizontal direction in the figure) in the gray shaded region for $8\text{ W} < P_{\text{pump}} < 10\text{ W}$ in Fig. 2(b), here we focus on $\varsigma_t$. Without a thermal lens (i.e., $P_{pump} = 0$ W and $f_{thermal} \to \infty$), $\varsigma_t$ is calculated to be about 4.5 rad, which is slightly less than $3\pi/2$. As $P_{pump}$ increases, $f_{thermal}$ becomes smaller, and $\varsigma_t$ gradually increases. When $\varsigma_t$ approaches $3\pi/2$ at $f_{thermal} \sim 20$ mm, the fourth-order transverse mode becomes degenerate with the fundamental mode, which degrades the output beam. We consider that this condition was satisfied when 8 W < $P_{pump}$ < 10 W, as shown by the first gray shaded region in Fig. 2. For higher $P_{pump}$, we observed two other gray shaded regions, where we consider that $\varsigma_t$ or $\varsigma_s$ approached $2\pi \times n/m$ with $m > 4$ and the transverse modes became degenerate again. This degradation of the power and the mode quality were observed in a narrower range of the pump power than in the first case, which is probably because lasing in the $m$-th order transverse mode with $m > 4$ was suppressed by spatial-mode mismatching to the transverse mode of the pump beam.

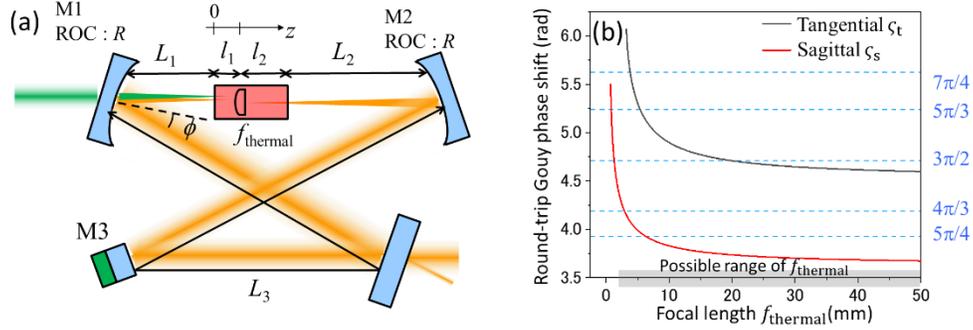

Fig. 4 Calculation to estimate the influence of the thermal lens on the Ti:sapphire laser. (a) A detailed view of the Ti:sapphire cavity. It is assumed that one thermal lens is formed in the crystal around the focal point of the pump laser. $f_{\text{thermal}}$ is the focal length of the thermal lens, $L_{1-3}$ and $l_{1,2}$ are the lengths of the parts shown by black arrows, $\phi$ is the incident angle on M1 and M2, and $R$ is the radius of curvature (ROC) of M1 and M2. (b) Calculated round-trip Gouy phase shift $\varsigma_{t,s}$ for the laser cavity. Black (red) curve indicates $\varsigma_t$ ($\varsigma_s$) in the tangential (sagittal) plane. When either of $\varsigma_{t,s}$ equals $2\pi \times n/m$ where $m$ and $n$ are integers, shown by the blue dashed lines for $m < 9$, the $m$-th order transverse mode in the Ti:sapphire cavity becomes degenerate with the fundamental modes, which degrades the output beam.

When the focal length decreases down to 3.2 mm, $(A_t + D_t)/2$ exceeds 1 in Eq. (2) and the laser cavity becomes unstable. This will be a fundamental limit of the present laser system. This situation appears when the pump power is increased to 50 W, implying that out power can be scaled up to 24 W if we can maintain the present laser configuration against other possible problems in the higher power operation, such as the thermal birefringence and higher-order thermal effects.

## 5. Summary

In this article, we have reported the development of a high-pump-power injection-locked CW Ti:sapphire laser. As a result of optimization of the waist diameter of a pump beam and an output coupler for a low-loss cavity configuration and shifting the focal length of a thermal lens via the temperature of the laser crystal holder, we achieved a 10 W output power with a slope efficiency of 51% with maintaining good transverse and longitudinal mode properties. We discussed a possible mechanism of the influence of the thermal lens on the cavity mode by using a simple model. Here we reported the laser performance at 780 nm, which is a resonant wavelength of rubidium atom. Similar performance will be expected at resonant wavelengths of other Alkali atoms or magic wavelengths of Alkali-earth atoms [28] considering the broad tunable wavelength range of Ti:sapphire lasers. High power and spectral purity, which were indicated in this article, would be useful for atom interferometers [5,14] or optical lattice clocks [4]. In the future, more detailed and quantitative modeling of this thermal lens mechanism may lead to more optimized designs. Another important step is to apply the thermal lens model to laser-diode pumped Ti:sapphire lasers, which would give us more compact lasers with high power and good mode properties [29].

### Acknowledgement

We would like to acknowledge Ken-ichi Ueda for useful comments and Koji Kamiyama for fruitful discussions.

### Disclosures

The authors declare no competing interests.